\documentclass[conference, a4paper]{IEEEtran}

\usepackage[pdftex]{graphicx}
\usepackage{amsmath}
\usepackage{algorithm}
\usepackage{algorithmic}

\usepackage{txfonts}

\IEEEoverridecommandlockouts
\renewcommand\IEEEkeywordsname{Keywords}

\begin{document}

\title{Using Conservative Estimation for Conditional Probability instead of Ignoring Infrequent Case}

\author{\IEEEauthorblockN{Masato Kikuchi\IEEEauthorrefmark{1},
Eiko Yamamoto\IEEEauthorrefmark{2},
Mitsuo Yoshida\IEEEauthorrefmark{1},
Masayuki Okabe\IEEEauthorrefmark{3},
Kyoji Umemura\IEEEauthorrefmark{1}}
\IEEEauthorblockA{\IEEEauthorrefmark{1}Department of Computer Science and Engineering\\
Toyohashi University of Technology,
Toyohashi 441-8580, Japan\\ \{m143313@edu,  yoshida@cs\}.tut.ac.jp, umemura@tut.jp}
\IEEEauthorblockA{\IEEEauthorrefmark{2}Department of Economics and Information\\
Gifu Shotoku Gakuen University,
Gifu 500-8288, Japan\\ eiko@gifu.shotoku.ac.jp}
\IEEEauthorblockA{\IEEEauthorrefmark{3}Faculty of Management and Information System\\
Prefectural University of Hiroshima,
Hiroshima 734-8558, Japan\\ okabe@pu-hiroshima.ac.jp}}

\IEEEpubid{\makebox[\columnwidth]{978--1--5090--1636--5/16/\$31.00~\copyright~2016 IEEE \hfill }
\hspace{\columnsep}\makebox[\columnwidth]{\hfill }}

\maketitle

\begin{abstract}
There are several estimators of conditional probability from observed frequencies of features.
In this paper, we propose using the lower limit of confidence interval on posterior distribution determined by the observed frequencies to ascertain conditional probability.
In our experiments, this method outperformed other popular estimators.
\end{abstract}

\begin{IEEEkeywords}
Conservative estimation, Conditional probability, Confidence interval, Lower limit.
\end{IEEEkeywords}

\IEEEpeerreviewmaketitle

\section{Introduction}

Estimating conditional probability from observed frequencies of features is the fundamental operation of natural language processing (NLP) and its practical application~\cite{Jimeno-Yepes:15}.
When we need to analyze the relationship between two features, we sometimes need to estimate conditional probability from the frequencies of the occurrence of these features.
For example, we may need to know what is the chance a document contains the word A under the condition that the document contains the word B.
One problem in estimating conditional probability could be low frequency.
i.e, in this case when only a few documents may contain word B.

When we need to estimate the probability from observed frequencies or samples, we usually use maximum likelihood estimator (MLE), which is an unbiased estimator, and asymptotically converges to true probability when we have infinite number of observations.
When only a few observations are available, we need to use various smoothing methods~\cite{Hazem:13,Chen:96}.
One of the classical smoothing methods is additive smoothing whose background is Bayesian framework.
In this framework, we assume prior distribution (Prior) of the features, and compute posterior distribution (Posterior) based on the observations, regardless of the number of observations.

When the number of observations is small, Posterior has large variance.
Therefore, we need to pay attention to deciding the estimated value from Posterior of the conditional probability.
When we choose the value that gives maximum probability in Posterior and we assume that Prior is uniform distribution, the value is same as MLE.
When we choose the expected value of the probability from Posterior, and assume that Prior is uniform distribution, the value is same as Laplace smoothing estimator.

In this paper, we focus our attention on this classical framework with a novel viewpoint.
We propose to form confidence interval of Posterior, using the lower limit of confidence interval for the estimator for judging the strength of the relationship between two features.
We conducted experiments where the true conditional probability is the true strength between features, and found that it outperformed other estimators.

\section{Related Work}

\IEEEpubidadjcol

Church and Hanks~\cite{Church:90}, proposed the concept of mutual information to measure the association between words.
This method solves low-frequency problem, because infrequent features cannot sum up to a large mutual information quantity.
However, if the true strength of the relationship between two features can be regarded as conditional probability, it may not be an appropriate measure to use because mutual information is symmetric towards two features, whereas conditional probability is asymmetric.

The Good-Turing estimator~\cite{Good:53,Gale:95} is a popular method to adjust frequency for the low frequency case.
This method needs to assume that the distribution of frequency obeys Zip's law.
We need to verify the distribution of features, and if it does not obey Zip's law, this method is not appropriate.

Conditional probability is more popular in the database field than NLP.
Apriori~\cite{Agrawal:94} is the most practically used method for finding the relationship between features, when the true strength between features is known to be conditional probability.
It uses MLE as the measure for the relationship.
To overcome the problem of low frequency, Apriori ignores the rare features using a threshold value, which is called minimum supports.
Although Apriori is efficient by ignoring low-frequency features, we sometimes need to compare a high frequency but low MLE value relationship with a low frequency but high MLE value relationship.
Apriori simply ignores relation of the low frequency but high MLE value relationship.

To overcome this problem, Predictive Apriori~\cite{Scheffer:05} is proposed, where Posterior of conditional probability of two features is computed based on Prior, which is decided by the distribution of the two features, and the expected value of conditional probability is used in stead of MLE.
We have examined Predictive Apriori, and found that the actual Prior and the Prior that shows the best performance are different.
This suggests that we need another parameter or viewpoints to decide the estimated value of strength of a relation.

When we need to control of the quality of product, we form the confidence interval of the probability of the chance that the product is defective.
Then, we use the upper limit of the confidence interval to estimate the ratio of defectiveness.
We use the upper limit because it causes more trouble if a defective product is judged as normal than if a normal product is judged as defective.
In the case of finding relationships, we need to be careful about the precision of the results found; a precision of 50\% is not considered satisfactory.
This suggests that it causes much more trouble if a false relationship is judged as true, than if a true relationship is judged as false.
In this case, the lower limit is natural choice for measuring the strength of a relationship.

Another issue in forming these confidence intervals is the low frequency of occurrence.
We need to form the confidence interval from less frequent features.
We find well-known approximation of confidence interval is not usable because it assumes that there are enough samples.
Though we may check enough samples to form the confidence interval by approximation formula for detecting defectiveness, we cannot increase the number of observation for detecting relationship.
Moreover, we have found that the so-called ``exact formula''~\cite{Clopper:34} of interval has considerable errors.
We have found that we need to numerically compute these lower limits of confidence.

\section{Posterior from Observed Frequencies}

\begin{figure}[tbp]
\centering
\includegraphics[scale=0.6]{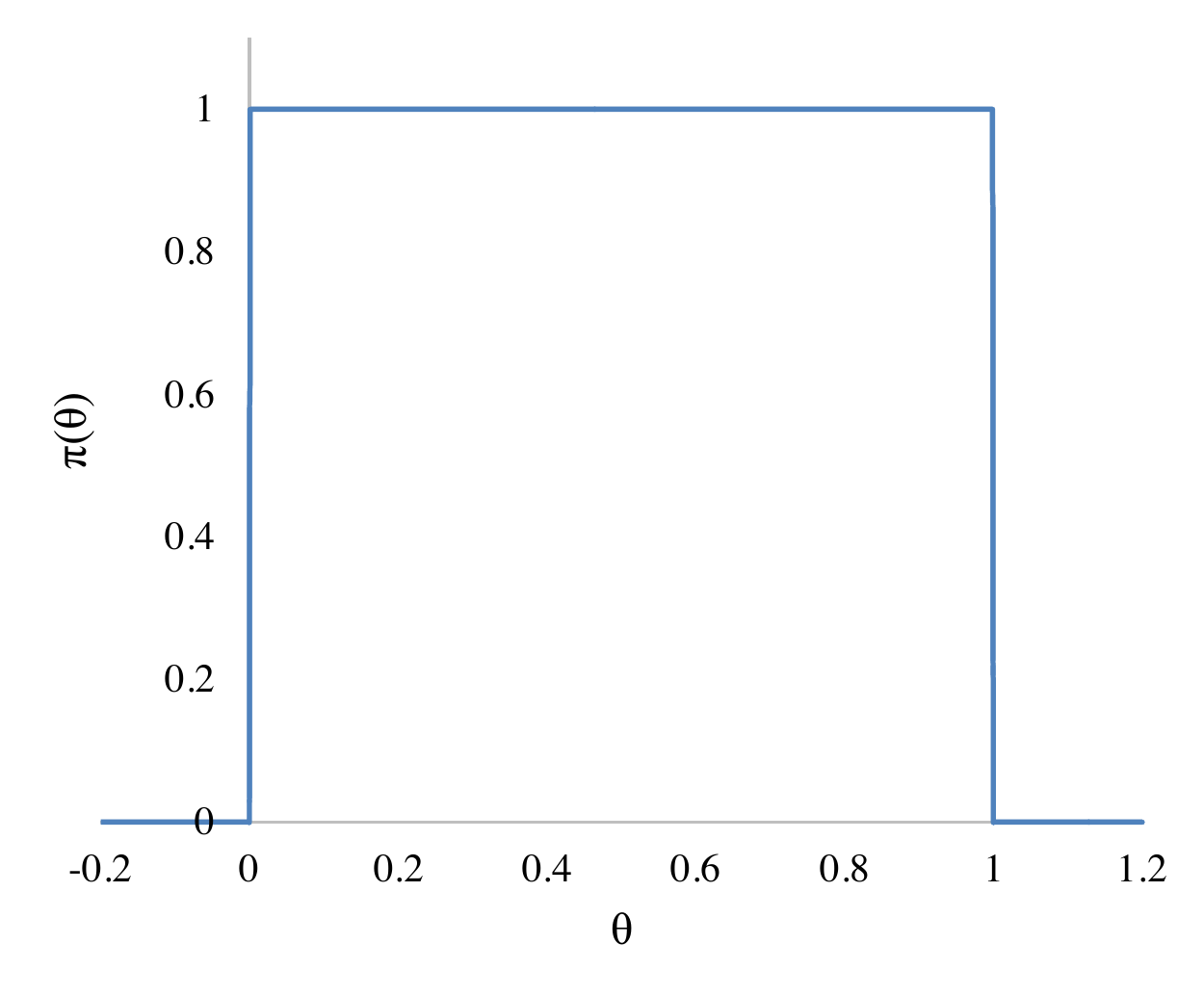}
\caption{
Uniform Distribution $\pi(\theta)$, which is $P(\Theta \mid N=0, X=0)$ that is used in experiment.
This Prior is usually used when we have no knowledge of $\theta$.
We have chosen this distribution because Apriori does not utilize the prior knowledge.
}
\label{fig:Uniform}
\end{figure}

\begin{figure}[tbp]
\centering
\includegraphics[scale=0.6]{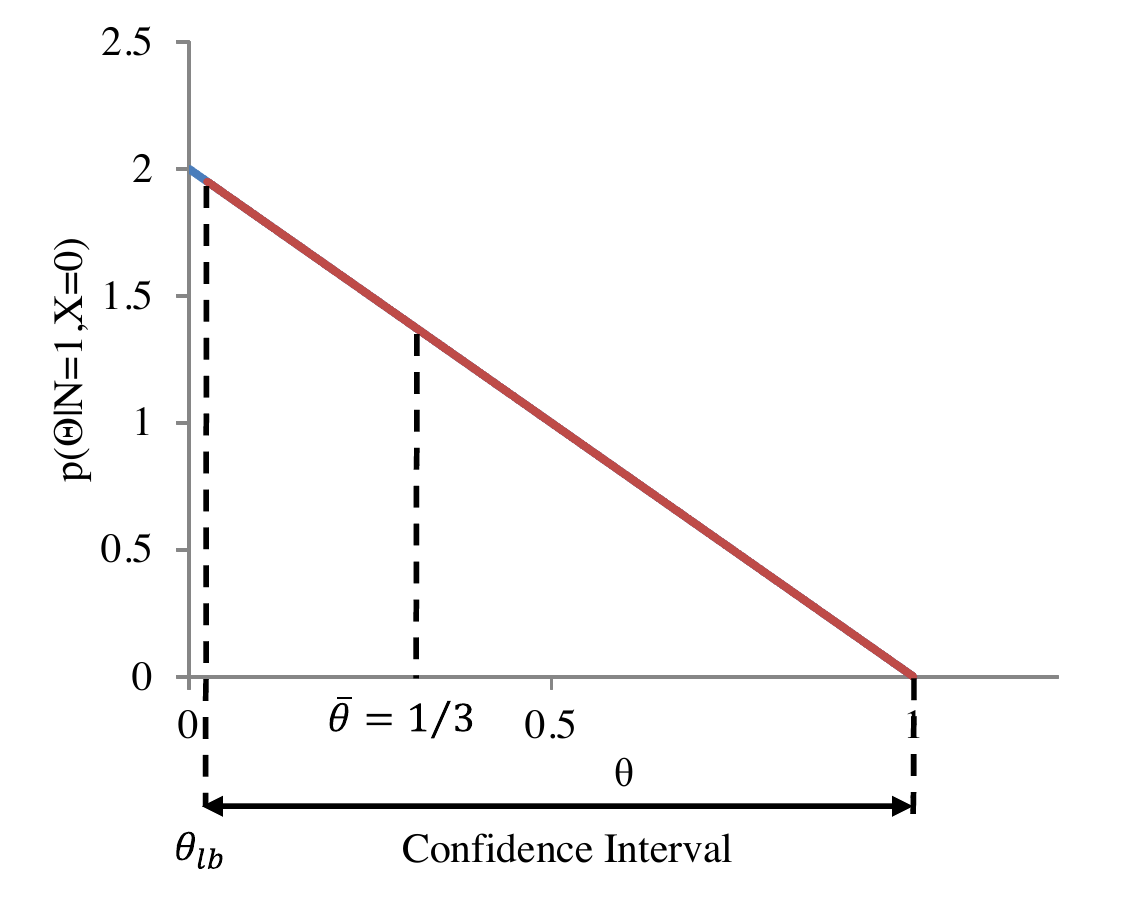}
\caption{
Posterior Distribution $P(\Theta \mid N=1, X=0)$, and its confidence interval [$\theta_{lb}$, 1].
}
\label{fig:Posterior1}
\end{figure}

\begin{figure}[tbp]
\centering
\includegraphics[scale=0.6]{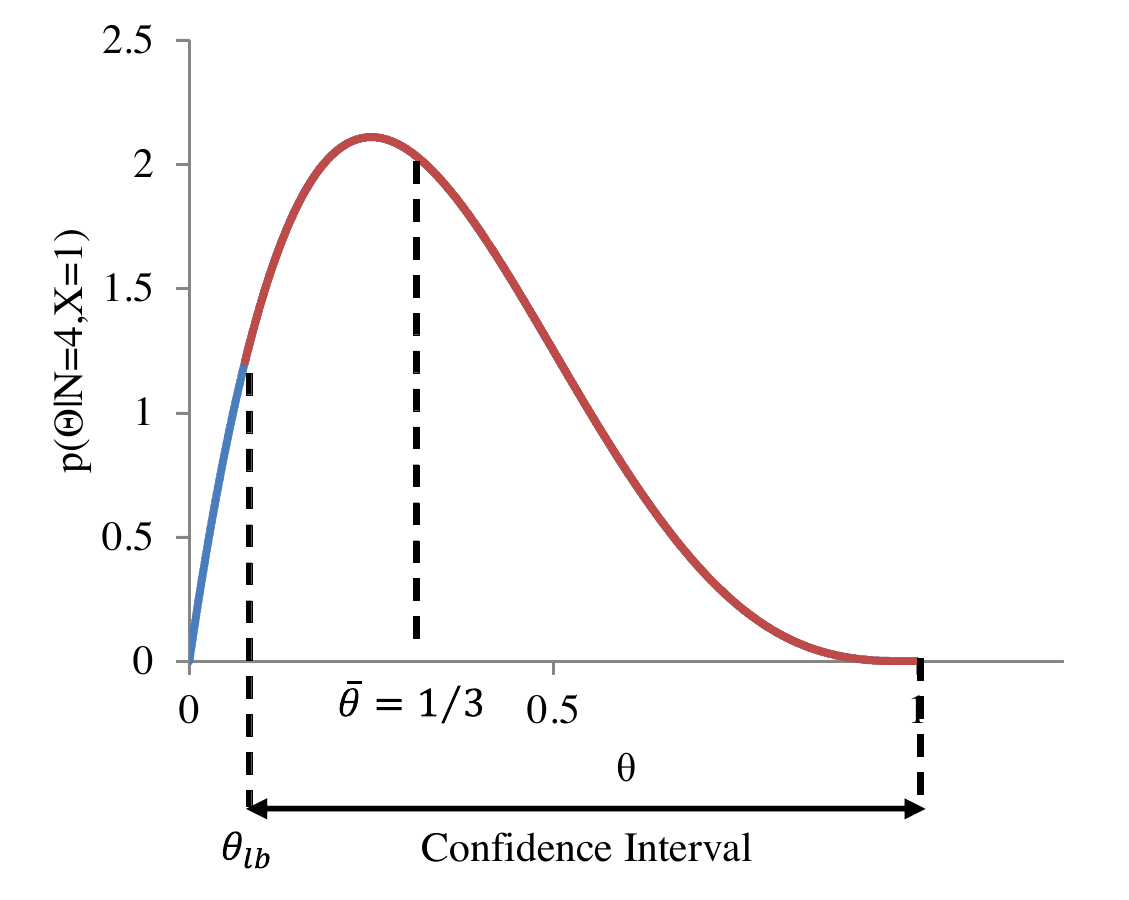}
\caption{
Posterior Distribution $P(\Theta \mid N=4, X=1)$, and its confidence interval [$\theta_{lb}$, 1].
It has same expected value as $P(\Theta \mid N=1, X=0)$, but larger $\theta_{lb}$, since its variance is smaller.
}
\label{fig:Posterior2}
\end{figure}

Let $\theta$ be the true value of conditional probability $P(A|B)$, which we need to estimate.
Let $n$ be the frequency of event B.
Let $x$ be the frequency of event A and B.
When we assume that Prior is uniform distribution $\pi(\theta)$, Posterior for $n$ and $x$ is as follows, where $L$ is normalization constant so that the integral from negative infinity to positive infinity should become 1.0.
\begin{align*}
	& P(\Theta \mid N=n,X=x) = & \\
	& \begin{cases}
		L \times \theta^{x}(1-\theta)^{n-x} & (0 < \theta < 1), \\
		0 & \textrm{Otherwise.}
	\end{cases} &
\end{align*}
$P(\Theta \mid N=0,X=0)$ is uniform distribution (Fig.~\ref{fig:Uniform}), and $P(\Theta \mid N=1,X=0)$, $P(\Theta \mid N=4,X=1)$ as shown in Fig.~\ref{fig:Posterior1}, and ~\ref{fig:Posterior2} respectively.

\section{Confidence Interval of Posterior}

The confidence intervals are in the range of $\theta$, where the probability that $\theta$ fall into this range is the value of the confidence level.
Although there may be many choices of intervals, the interval that we form is [$\theta_{lb}$, 1], where the  lower bound $\theta_{lb}$ is defined as follows:
\[
	P(\Theta > \theta_{lb} \mid N,X)=\alpha,
\]
where $\alpha$ is confidence level.
Please note that even for $P(\Theta \mid N=1,X=0)$, $\theta_{lb}$ is positive.
Although usual confidence interval of $\theta$ contains the value 0, the proposed confidence interval will never contains the value 0.
Therefore $\theta_{lb}$ can be regarded as a conservative smoothing value.
We have chosen confidence level $\alpha$ by considering the required precision of estimation.

The value $\theta_{lb}$ is determined by $n$, $x$, $\alpha$ and Prior.
Therefore, we can have a table of $\theta_{lb}$ before judging the strength of the relation.
For making the check experiment easier the table of $\theta_{lb}$ by $n$, $x$ for $\alpha$ = 0.99, using uniform distribution as Prior is available on the Web, whose URL is ``http://www.ss.cs.tut.ac.jp/CI-Laplace''.
For $n < 7$, the value is shown in the Appendix.

\section{Experimental Setting}

\begin{algorithm}[tbp]
\caption{Generation algorithm of the synthetic dataset}
\label{alg:alg1}
\begin{algorithmic}
\STATE $D^* \coloneqq \phi$;\ $k \coloneqq 0$;
\WHILE{$k<1000$}
	\STATE $j \coloneqq 0$;\ $t_k \coloneqq \phi$;
	\WHILE{$j<2$}
		\STATE extract $\langle S_l,C_m \rangle$ at random from $R$;
		\STATE $t_k \coloneqq t_k \cup S_l \cup C_m$;
		\STATE $j \coloneqq j+1$
	\ENDWHILE
	\STATE $D^* \coloneqq D^* \cup \{ t_k \}$;
	\STATE $k \coloneqq k+1$
\ENDWHILE
\end{algorithmic}
\end{algorithm}

\begin{figure}[tbp]
\centering
\caption{
Algorithm for preparing dataset.
R is hierarchical relation, and we have chosen actual names of prefectures (corresponds to state) and cities.
}
\label{fig:Alg1}
\end{figure}

\begin{table}[tbp]
\centering
\caption{Statistics of the Synthesized Data}
\begin{tabular}{| l | r |} \hline
Number of transactions & 1,000 \\ \hline
Kinds of pairs of candidate pairs & 4,469 \\ \hline
Number of occurrences of candidates pairs & 5,934 \\ \hline
Kinds of right pairs & 975 \\ \hline
Number of occurrences of right pairs & 2,000 \\ \hline
\end{tabular}
\label{tab:Dataset}
\end{table}

We synthesized the dataset to clearly compare the estimators. In this synthesis, the task is to separate hierarchical relationships from their mixed data.
Let R be the actual hierarchical relationship of prefectures (or states) and cities.
For example, R could be $\{ \langle S_1, C_1 \rangle , \langle S_1, C_2 \rangle , \langle S_1, C_3 \rangle , \langle S_2, C_4 \rangle , \langle S_2, C_5 \rangle , \langle S_3, C_5 \rangle \}$ , where $S_i$ and $C_i$ are the  name of prefectures, and cities, respectively.
We randomly selected two relationships to synthesize the dataset.
For example, the synthesized data could be $\{\{C_5, C_1, S_2, S_1\}, \{C_3, C_4, S_1, S_2\}, ...\}$.
As different cities may belong to the same prefecture (or state), there is nothing wrong if the prefecture name of one city appears in the data of another city.
Therefore, the correct measure for estimating the relationship between a city and its prefecture is the conditional probability of the prefecture under the condition that the city name appears.
The algorithm and statistics of the synthesized data are shown in Fig.~\ref{fig:Alg1} and TABLE~\ref{tab:Dataset} respectively.

For the generated dataset, we first estimate the conditional probabilities of all the pairs of city/prefecture names.
We then rank the list of name pairs from the largest estimation value to the smallest.
Lastly, we assess whether each pair exists in R and compute the recall and precision using the pairs in order from the first to the current one.
The obtained list could be ($\langle S_1, C_1 \rangle , \langle S_3, C_5 \rangle , \langle S_1, C_5 \rangle , \langle S_2, C_4 \rangle , ...$).
In this example, only the $\langle S_1, C_5 \rangle$ is not in R, and there are five relationships in R.
Then the values of recall at the $i_{th}$ pair are (1/5, 2/5, 2/5, 3/5, ...).

We observe the estimator performance by plotting the recall by rank.
In this plot, we can also see the precision of each rank as the slope of the line passing through the origin and each plotted point.
The line of the best estimator will appear above the other lines.
The proposed method has a single parameter, $\alpha$ (confidence level).
In our experiment, we choose $\alpha=0.99$ as the desired precision for high (small) rank.

\subsection{Comparison with MLE (Apriori)}

\begin{figure}[tbp]
\centering
\includegraphics[scale=0.55]{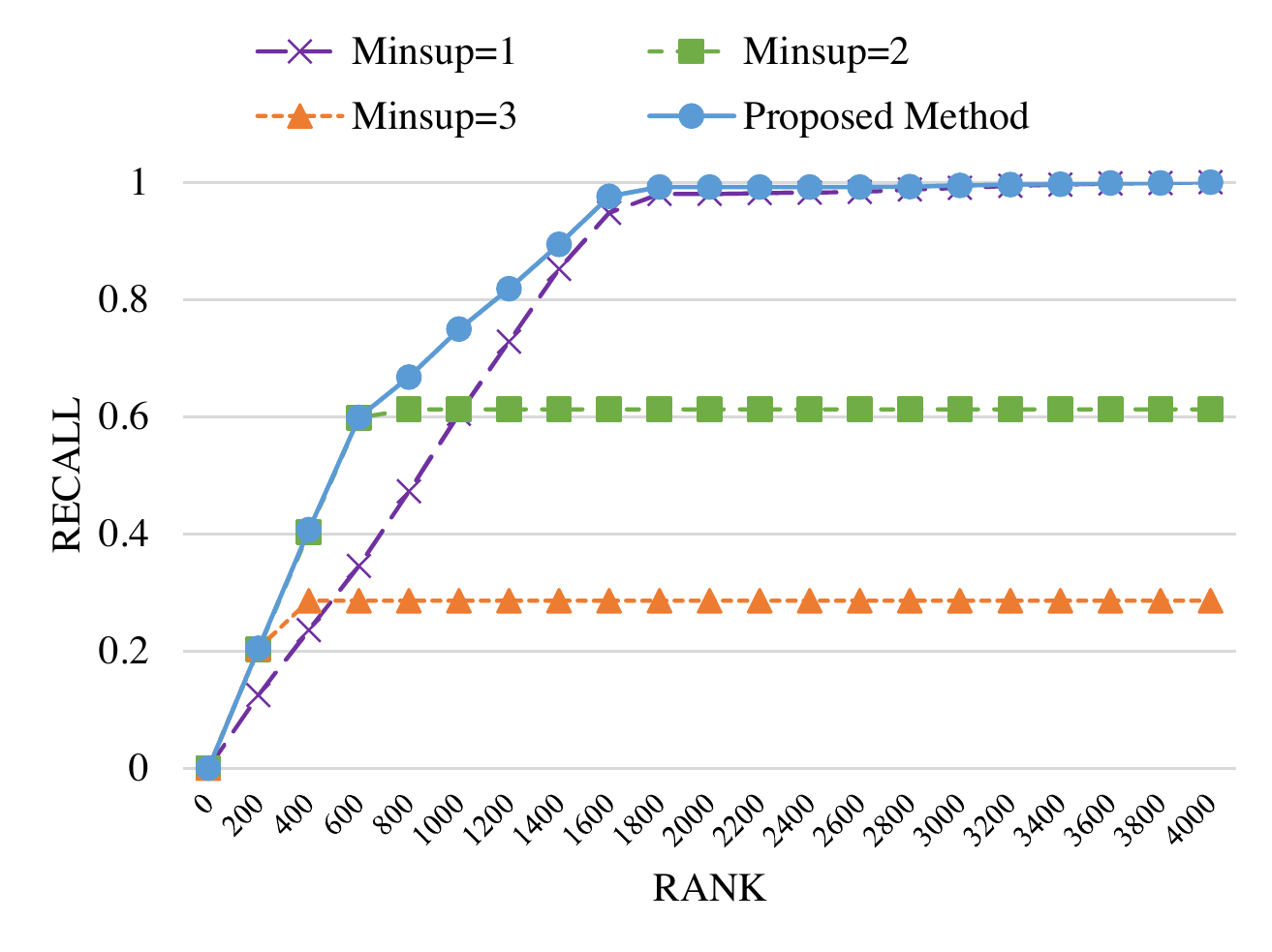}
\caption{
Recall rate by $\hat{\theta}$ and $\theta_{lb}$.
Minsup means minimum supports.
Minsup=1 corresponds to maximum likelihood estimator $\hat{\theta}$.
Setting the appropriate value for minimum support benefits of the high (small) rank range with cost of the low (large) rank range.
The proposed method $\theta_{lb}$ always better than maximum likelihood estimator $\hat{\theta}$ with any Minsup value.
}
\label{fig:Apriori}
\end{figure}

It is a common practice to ignore the low-frequency case.
Apriori~\cite{Agrawal:94} does this, and calls the minimum frequency as minimum support (minsup for short).
We conducted an experiment changing the minimum support.
When minimum support is 1, it is equivalent to use ordinal MLE $\hat{\theta}$ as estimator.
As shown in Fig.~\ref{fig:Apriori}, setting the appropriate value for minimum support benefits of the high (small) rank range with cost of the low (large) rank range.
This result is obtained by completely ignoring the doubtful relation due to low frequency.

In the case of $\theta_{lb}$, we can obtain a similar improvement in the high rank range by discounting the probability of low frequency.
In our case, low frequency cases still have some positive value and may have chance of being included in the output.
Therefore, we lose nothing in the low rank range, as shown in Fig.~\ref{fig:Apriori}.

Both Apriori and $\theta_{lb}$ have one parameter to choose, minimum support and confidence level.
Our result suggests $\theta_{lb}$ with the appropriate confidence level, always outperforms Apriori, regardless with the value of its minimum support.

\subsection{Comparison with expected value using Posterior and Prior}

\begin{figure}[tbp]
\centering
\includegraphics[scale=0.65]{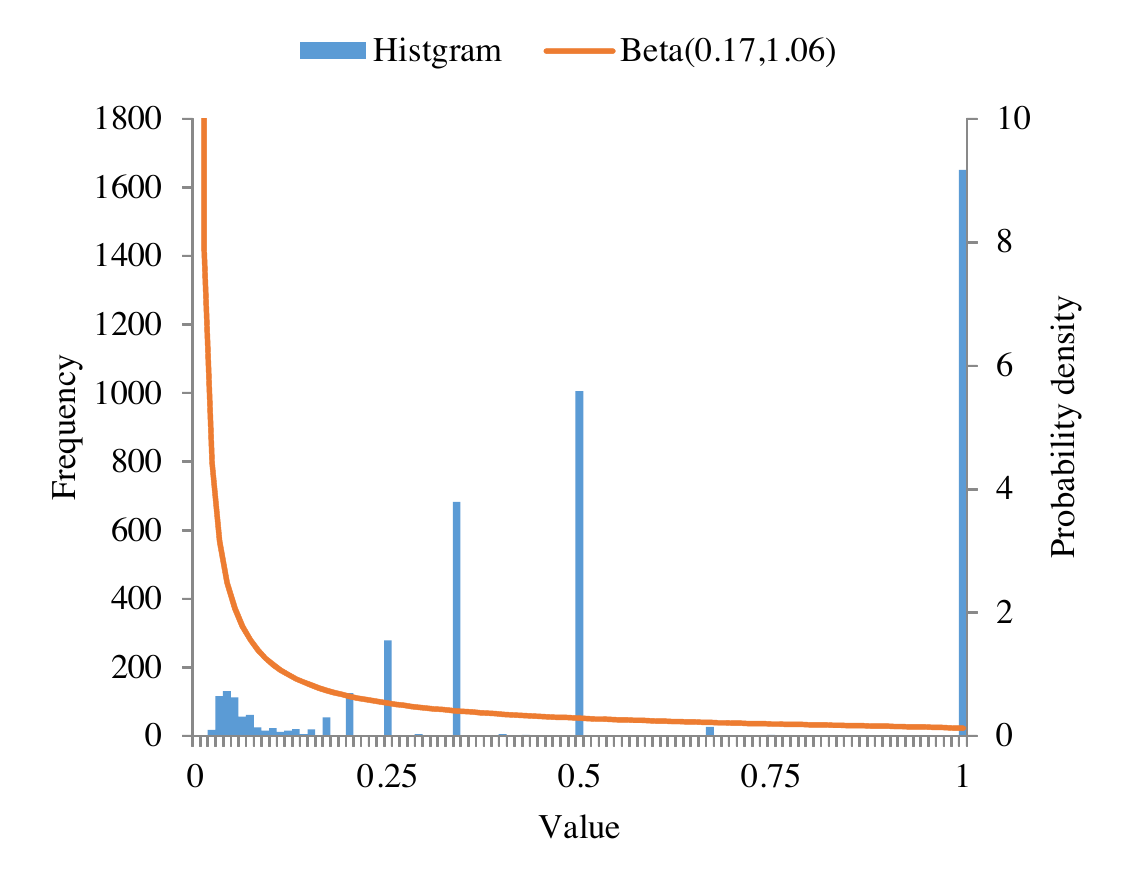}
\caption{
Prior Distribution of $\theta$ in the dataset for all name pairs (blue), and Prior distribution used in the experiment (red).
In determining the Prior distribution, note that we ignore low-frequency data, e.g., 1/1, 1/2, 1/3, 2/3, 1/4, 3/4, 1/5, 2/5, 3/5 and 4/5.
}
\label{fig:Prior}
\end{figure}

\begin{figure}[tbp]
\centering
\includegraphics[scale=0.55]{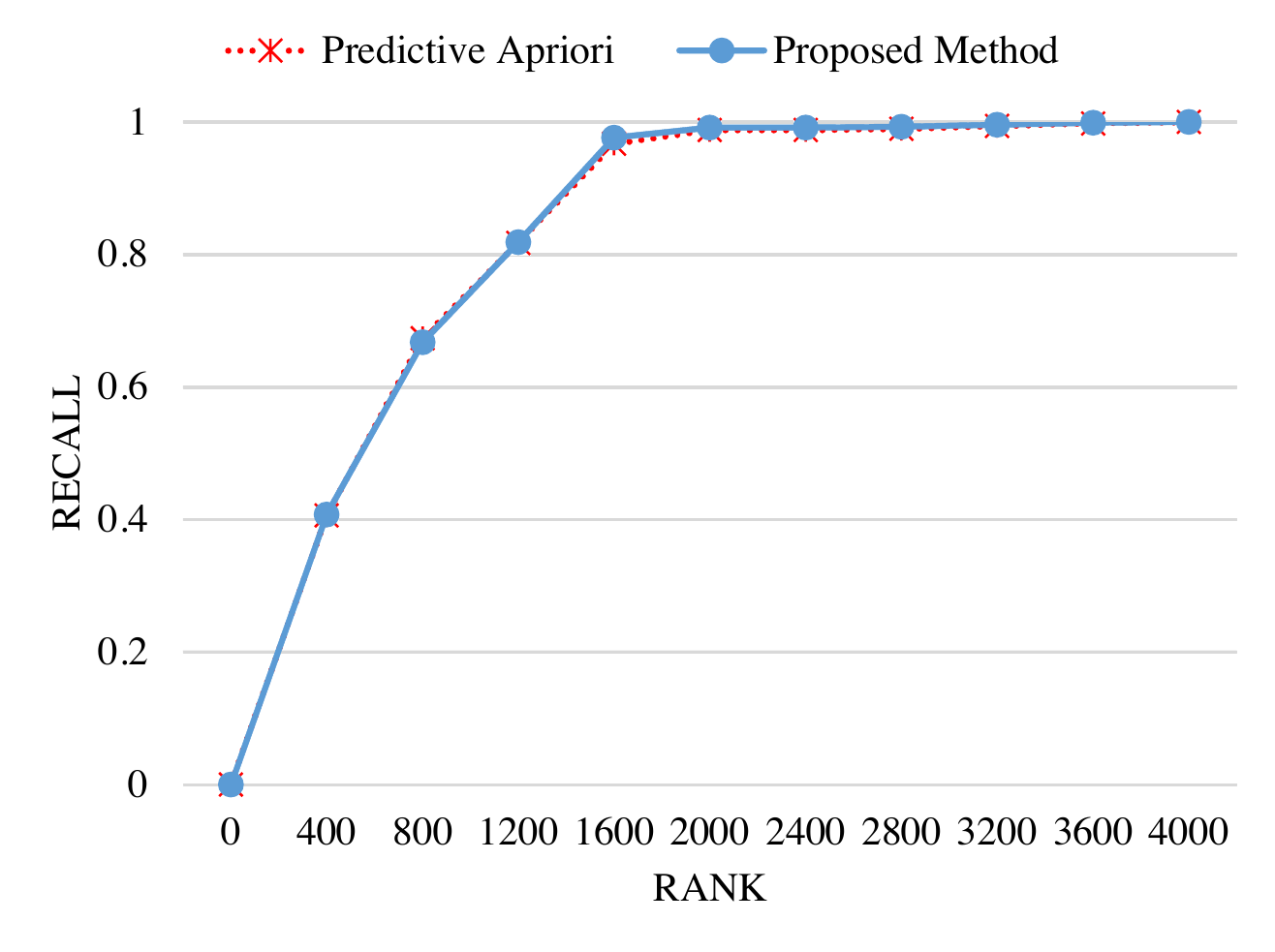}
\caption{Recall rate by $\bar{\theta_p}$ and $\theta_{lb}$.
Predictive Apriori $\bar{\theta_p}$ uses the beta distribution as the Prior distribution.
The parameters of the beta distribution are determined by examining the dataset.
Although we get approximately the same results, Predictive Apriori needs to estimate the Prior distribution, which is not always straightforward to do.
}
\label{fig:Predictive}
\end{figure}

The Predictive Apriori algorithm~\cite{Scheffer:05} uses $\bar{\theta_p}$, the expected value of $\theta$ in the Posterior distribution.
Predictive Apriori needs the Prior of $\theta$ to compute the Posterior.
Fig.~\ref{fig:Prior} shows the histogram (in blue) of $\theta$ for all name pairs.
Spikes can be observed at 1/1, 1/2, and 1/3, corresponding to the low-frequency pairs.
Although Predictive Apriori usually uses this histogram of data as its Prior distribution, assuming that the Prior distribution has this shape would be wrong.
There is no reason that $\theta$ is likely to be a particular number, such as 1/2.
Generally speaking, the estimation of the Prior distribution is not an easy task.
This estimation requires a multitude of considerations in an actual situation.

By observing the histogram, we choose to ignore the data of 0, 1/1, 1/2, 1/3, 2/3, 1/4, 3/4, 1/5, 2/5, 3/5, 4/5.
The modified the histogram is then smoothed as a ?½?½ distribution, allowing the remaining case to be observed in the resulting distribution. The beta distribution has two parameters.
We usually determine these on the basis of observed mean and variance. :
\[
	\beta(a,b)=\frac{\theta^{a-1}(1-\theta)^{b-1}}{ \int_0^1 t^{a-1}(1-t)^{1-b} dt} \quad (0 < \theta < 1).
\]
The parameters are determined by inspecting the dataset. As a result, we use?½?½(0.17, 1.06), which gives the red curve shown in Fig.~\ref{fig:Predictive}.

Please note that, this Prior distribution has a relatively large probability for a relative small value of $\theta$.
Using this distribution, $\bar{\theta_p}$ also behaves like the proposed method.
When we have only a few name occurrences, the shape of Posterior is close to that of Prior.
Thus, $\bar{\theta_p}$ becomes less than $\hat{\theta}$.
For the evaluation of $\bar{\theta_p}$, we obtained almost the same curve with the proposed method.
Although we can get similar results, estimating the Prior distribution is a more complex operation and less intuitive than the proposed method.
If we choose the beta distribution as the Prior distribution, the number of parameters are less than histograms.
Still there are two parameters $a$ and $b$, whereas $\theta_{lb}$ has only one parameter $\alpha$ (confidence level).

\section{Conclusion}

We have proposed to use $\theta_{lb}$ instead of $\hat{\theta}$ or $\bar{\theta}$.
We chose the interval as [$\theta_{lb}$, 1], and the confidence level $\alpha$ as the required precision of the output.
Using a synthesized dataset, we have found that $\theta_{lb}$, outperformed $\hat{\theta}$, with or without minimum support.
We have also compared $\theta_{lb}$ and $\bar{\theta_p}$.
Although $\theta_{lb}$ and $\bar{\theta_p}$ show almost the same results, we need to know the Prior distribution of $\theta$ to use $\bar{\theta_p}$, and determing the Prior distribution is not always an easy task.

Even in the case where Prior distribution is well known, and $\bar{\theta_p}$ seems appropriate, there is a possibility of forming a confidence interval for this estimator.
Even when the distribution of features obeys Zip's law and the Good-Turing estimator seems appropriate, there again lies a possibility of forming a confidence interval.
Our proposal is to consider how the estimator is treated, and to provide an additional viewpoint on selecting the estimator to use.

\section*{Acknowledgment}

This work was supported by 2015 Gifu Shotoku Gakuen University Research Grant.

\bibliography{ICAICTA2016}
\bibliographystyle{IEEEtran}

\section*{Appendix}

\begin{table}[htbp]
\centering
\caption{Table of Various Estimators, using the Uniform Distribution as the Prior Distribution of $\theta$}
\begin{tabular}{| r | r | r | r | r |} \hline
\multicolumn{1}{| c |}{$n$} & \multicolumn{1}{ c |}{$x$} & \multicolumn{1}{ c |}{$\hat{\theta}$} & \multicolumn{1}{ c |}{$\bar{\theta}$} & \multicolumn{1}{ c |}{$\theta_{lb}$} \\ \hline
1 & 0 & 0.00000 & 0.33333 & 0.00501 \\
1 & 1 & 1.00000 & 0.66667 & 0.10000 \\
2 & 0 & 0.00000 & 0.25000 & 0.00334 \\
2 & 1 & 0.50000 & 0.50000 & 0.05890 \\
2 & 2 & 1.00000 & 0.75000 & 0.21544 \\
3 & 0 & 0.00000 & 0.20000 & 0.00251 \\
3 & 1 & 0.33333 & 0.40000 & 0.04200 \\
3 & 2 & 0.66667 & 0.60000 & 0.14087 \\
3 & 3 & 1.00000 & 0.80000 & 0.31623 \\
4 & 0 & 0.00000 & 0.16667 & 0.00201 \\
4 & 1 & 0.25000 & 0.33333 & 0.03268 \\
4 & 2 & 0.50000 & 0.50000 & 0.10564 \\
4 & 3 & 0.75000 & 0.66667 & 0.22207 \\
4 & 4 & 1.00000 & 0.83333 & 0.39811 \\
5 & 0 & 0.00000 & 0.14286 & 0.00167 \\
5 & 1 & 0.20000 & 0.28571 & 0.02676 \\
5 & 2 & 0.40000 & 0.42857 & 0.08473 \\
5 & 3 & 0.60000 & 0.57143 & 0.17307 \\
5 & 4 & 0.80000 & 0.71429 & 0.29431 \\
5 & 5 & 1.00000 & 0.85714 & 0.46416 \\
6 & 0 & 0.00000 & 0.12500 & 0.00144 \\
6 & 1 & 0.16667 & 0.25000 & 0.02267 \\
6 & 2 & 0.33333 & 0.37500 & 0.07080 \\
6 & 3 & 0.50000 & 0.50000 & 0.14227 \\
6 & 4 & 0.66667 & 0.62500 & 0.23632 \\
6 & 5 & 0.83333 & 0.75000 & 0.35664 \\
6 & 6 & 1.00000 & 0.87500 & 0.51795 \\ \hline
\end{tabular}
\label{tab:Estimators}
\end{table}

\end{document}